\documentclass[conference]{IEEEtran}
\IEEEoverridecommandlockouts
\usepackage{cite}
\usepackage{amsmath,amssymb,amsfonts}
\usepackage{algorithmic}
\usepackage{graphicx}
\usepackage{textcomp}
\usepackage{xcolor}
\usepackage{booktabs}
\usepackage{tabularx}
\usepackage{subcaption}
\usepackage{multirow}
\usepackage{url}
\usepackage{makecell}
\usepackage{fancyvrb}
\def\BibTeX{{\rm B\kern-.05em{\sc i\kern-.025em b}\kern-.08em
    T\kern-.1667em\lower.7ex\hbox{E}\kern-.125emX}}
\begin{document}

\title{Multilingual AI-Driven Password Strength Estimation with Similarity-Based Detection\\
}

\author{\IEEEauthorblockN{\textsuperscript{} Nikitha M. Palaniappan}
\IEEEauthorblockA{\textit{School of Computer Science and Electronic Engineering} \\
\textit{Queen Mary University of London}\\
London, UK \\
nikipalani@gmail.com}
\and
\IEEEauthorblockN{\textsuperscript{} Ying He}
\IEEEauthorblockA{\textit{School of Computer Science and Electronic Engineering} \\
\textit{Queen Mary University of London}\\
London, UK \\
yinghe@qmul.ac.uk}
}

\maketitle

\begin{abstract}
Considering the rise of cyberattacks incidents worldwide, the need to ensure stronger passwords is necessary. Developing a password strength meter (PSM) can help users create stronger passwords when creating an account on an online platform. This research aimed to explore whether incorporating a non-English training dataset (specifically Indian) can improve the performance of a PSM. Findings show that PSMs can be improved by utilising learning of words from other languages. Another contribution of the research was to compare and provide an analysis of AI generated data (specifically by ChatGPT) and PassGAN (existing state-of-the-art model), proving that PassGAN-like tools may no longer be needed as the performance is higher using AI generated data. To further strengthen detection, a Jaro similarity-based matching mechanism was incorporated, enabling the classification of passwords that are highly similar to known weak passwords - this addresses limitations of direct matching techniques used in prior work. A final novel contribution is on developing a PSM tailored for Indian passwords, which has not been developed previously - this resulted in a near-perfect matching accuracy using a Jaro function value of $0.5$. Although performance improvements were constrained by limited data and training, results suggest that using the ChatGPT dataset is a viable and effective strategy for developing secure, language-aware password strength meters. 


\end{abstract}


\section{Introduction}
Password security remains a serious concern in the current information systems as user-chosen passwords show predictable patterns despite decades of policy enforcement and user training. Traditional password strength assessment mechanisms, such as rule-based checks \cite{gusaas} and entropy calculations \cite{gusaas}, \cite {abdullahi}, have shown to be insufficient against large-scale password guessing attacks, which are mostly driven by leaked datasets and significant advances in computational power. As a result, recent research has increasingly focused on data-driven and deep learning-based approaches to model real-world password behaviour to improve both password guessing and strength evaluation \cite{hitaj}. This study builds upon the current research, with a particular emphasis on generative models, multilingual password analysis, and similarity-based evaluation.

Early hybrid password strength meters (PSM) approaches attempted to bridge the gap between rule-based and learning-based methods. GENPass, proposed by authors in \cite{liu}, combined Probabilistic Context-Free Grammar (PCFG) techniques with adversarial deep learning to generate password guesses. The approach first learns structural rules from real-world password datasets and then apply adversarial generation to produce realistic yet diverse password candidates. While this hybrid approach demonstrated the benefits of structure-aware generation and achieved a $23$\% match rate using a generated list of $10^{12}$ passwords - its complexity and reliance on large-scale generation limited its practicality. Nevertheless, GENPass is conceptually significant as it illustrates how adversarial learning can enhance traditional rule-based systems - an idea closely aligned with later GAN-based methods.

Moving beyond grammar-based representations, the study in \cite{pasquini} introduced a novel representation learning approach that reframed password guessing as a semantic learning problem. Instead of predicting passwords character by character, their model learned dense vector embeddings using neural networks and autoencoders. This enabled semantic clustering of passwords and improved generalisation to unseen patterns, achieving a notable $51.8$\% match rate with $10^{10}$ generated guesses. This work addressed key limitations of sequence-based models, such as exposure bias, where models overfit frequent character transitions and produce repetitive outputs. However, the approach is data-intensive and raises concerns about robustness across diverse datasets. However, the acceptable performance of representation learning establishes an important benchmark and motivates future exploration of combining semantic understanding with generative modelling.

Sequence-based neural architectures have also been used in password research such as those investigated in  \cite{melicher} that framed password “guessability” as a predictive task using Long Short-Term Memory (LSTM) networks trained on large, leaked datasets such as RockYou. Their model achieved a higher match rate of $70$\% using $10^{15}$ guesses, showing the ability of recurrent neural networks in capturing sequential character dependencies. However, high computational requirements and overfitting concerns limit such brute-force deep learning approaches.

The introduction of Generative Adversarial Networks (GANs) marked a significant shift towards fully data-driven password generation. The work by \cite{hitaj} proposed PassGAN, applying GANs to password guessing without any designed rules or heuristics. It was trained on leaked password datasets and evaluated against the LinkedIn breach - PassGAN achieved a $30.6$\% match rate and demonstrated strong synergy when combined with traditional password lists such as Best64. While GANs require careful tuning and large datasets, PassGAN’s unsupervised approach allows easier development of modern password strength assessment and guessing systems. Subsequent implementations such as those by \cite{D3vil0p3r}, further developed this approach - this repository serves as the pretrained baseline for this research.

A range of other PSMs have also been developed in practice, each adopting different trade-offs between accuracy, usability, and computational cost. Tools such as zxcvbn \cite{dropbox} and others \cite{wheeler_lowe}, \cite{wolf} leverage leaked datasets, entropy measures, and pattern matching to provide realistic strength estimates and user feedback, while rest such as OWASP’s Password Policy Project \cite{owasp_1}, \cite{owasp_2} prioritise policy compliance and lightweight integration. Although these tools are effective in certain specific contexts, these lack generative capabilities of advanced similarity-based evaluation; thereby limiting their ability to model real-world attack scenarios.

In summary, these studies highlight a progression from rule-based and sequential models towards 
generative, data-driven approaches that better capture real-world password distributions. However, existing work largely focuses on English-language datasets and exact-match evaluation metrics. This research addresses these gaps by investigating whether large language models powered by generative AI, such as ChatGPT \cite{chatgpt}, can be used as an alternative to GAN-based systems for password strength assessment and generation, with a particular focus on multilingual performance. In addition, no prior work has developed or evaluated an Indian-language-specific password model, representing a novelty of this study. Finally, the incorporation of similarity measures such as the Jaro function introduces a more specialised evaluation framework, reflecting realistic attack success beyond exact matches.

\section{Summary of the Methods}

This research used a data-driven password generation approach based on ChatGPT instead of PassGAN. The aim was to investigate whether passwords generated by a large language generative AI model could achieve good matching performance against real leaked password datasets, and whether including multiple languages would improve performance. The summary of the method was modified from the original GAN-based design that simplifies the implementation and computational time requirements.

\subsection{Password Dataset Generation}

The first step was to create password training word lists using ChatGPT. Three types of password datasets were generated: a Indian password dataset, an English password dataset, and a mixed dataset containing both Indian and English words.

All generated passwords followed the same structure in order to maintain consistency. Each password contained $8$-$10$ characters and included at least: one uppercase letter, one lowercase letter, one number, and one symbol. ChatGPT prompting required the generation to contain meaningful parts of real words rather than random characters, so that the passwords appeared realistic. The Indian dataset included cultural references such as Indian names, foods, and religious words. The English dataset included common English words and patterns. The mixed dataset combined English and Indian word fragments to simulate multilingual password behaviour. ChatGPT generated $19,998$ passwords for testing: $6,666$ English, $6,666$ Indian, and $6,666$ mixed (there was a limit on how much ChatGPT would generate, as there is a warning on fraudulent usage). Usage of ChatGPT to generate the passwords is appropriate here because it can produce realistic word-based patterns that are similar to human-created passwords. Samples of these generated passwords are shown in Table \ref{train_passwords} below.

\begin{table}[h]
\centering
\caption{ChatGPT Generated Sample Passwords}
\label{train_passwords}
\setlength{\extrarowheight}{5pt}
\begin{center}
\begin{tabular}{|p{1.8cm}|p{1.8cm}|p{1.8cm}|}
\hline
\multicolumn{1}{|c|}{\textbf{English}} & 
\multicolumn{1}{|c|}{\textbf{Indian}} & 
\multicolumn{1}{|c|}{\textbf{Hybrid}} \\ \hline
\verb|^|man*gB9, \newline W7home\verb|^|irs, \newline john9wC\verb|^| & raja\&Uh0,\newline dosa6\&-pT,\newline \#punem)C3 & 2=Jayearn,\newline tramAL\#I3,\newline \verb|_|chaigP4 \\ \hline
\end{tabular}
\end{center}
\end{table}


To test, two real leaked password datasets were used, one Indian and one English. An Indian leaked password data set \cite{indian_test_dataset} containing approximately $9,300$ passwords and an English leaked LinkedIn password data set \cite{D3vil0p3r} of $15,000$ passwords were used. The datasets were filtered so that passwords were between $8$ and $10$ characters long, in order to match the structure of the generated passwords resulting in a final test dataset of $7,675$ passwords for Indian and $11,356$ passwords for English. 

These datasets were treated as real-world weak passwords, because they had already been leaked. Samples of these passwords are shown in Table \ref{baseline_test_passwords}. 


\begin{table}[h]
\centering
\renewcommand{\arraystretch}{1.5} 
\caption{Baseline and Test Sample Passwords}
\label{baseline_test_passwords}
\begin{tabular}{|p{1.8cm}|p{1.8cm}|p{1.8cm}|} 
\hline
\multicolumn{1}{|c|}{\rule{0pt}{17pt}{\makecell{\textbf{PassGAN} \\ \textbf{(Baseline)}}}} & 
\multicolumn{1}{|c|}{\rule{0pt}{17pt}{\makecell{\textbf{LinkedIn} \\ \textbf{(English Test)}}}} & 
\multicolumn{1}{|c|}{\rule{0pt}{17pt}{\makecell{\textbf{zxcv32} \\ \textbf{(Indian Test)}}}} \\ \hline
loveloeod, \newline andababa, lukeloo3  & jeff102030, chocolate!, works20979 & Kohli@123, India@05, Airtel@002  \\ \hline
\end{tabular}
\end{table}

\subsection{Password Matching Method}

In order to mimic attacks in the real world, instead of exact matching, a similarity-based matching approach was used for testing the PSM. The Jaro similarity function \cite{jaro} was applied to compare generated passwords with real leaked passwords. The Jaro function measures how similar two strings are and produces a value between $0$ and $1$. The Jaro distance $d$ between two strings $s_1$ and $s_2$ is calculated as:

\begin{equation}
d(s_1,s_2) = \frac{1}{3} \left( \frac{m}{|s_1|} + \frac{m}{|s_2|} + \frac{m - t}{m} \right)
\end{equation}

where:
\begin{itemize}
    \item $m$ is the number of matching characters;
    \item $|s_i|$ is the length of string $s_i$;
    \item $t$ is the number of transpositions (matching characters not in right order divided by two).
\end{itemize}

Figure \ref{jaro_examples} shows some examples of the Jaro function usage. 

\begin{figure}[htbp]
\centering
\begin{minipage}{0.4\textwidth}
    \begin{Verbatim}[frame=single, fontfamily=courier, fontsize=\footnotesize, framesep=5mm]
>>> jaro('Brian', 'Jesus')
0.0
>>> jaro('Thorkel', 'Thorgier')
0.779761...
>>> jaro('bunty', 'bunti')
0.866666...
    \end{Verbatim}
\end{minipage}
\caption{Examples to illustrate how Jaro (string1, string2) function can be used}
\label{jaro_examples}
\end{figure}


A similarity threshold of $0.5$ was used. Similarity matching was used because attackers often guess passwords that are similar but not identical to real passwords. Therefore, similarity provides a more realistic measure than exact matching i.e. it would be a better measure of the true performance of the PSM. Here, if the similarity was more than $0.5$, passwords were considered a match, and if the similarity was less than $0.5$, passwords were considered different. This Jaro threshold was used as it provides a more realistic matching scenario, otherwise even if there is only a single letter match, the two passwords would be considered to be matched with a low value. Similarly, with a very high value, matching would only work with near-perfect cases, something that hackers generally will not have. Moreover, our Indian password experiment (discussed later) revealed this threshold to be appropriately used. 
\subsection{Evaluation Approach}

The performance of the system was measured using matching accuracy. Matching accuracy represents how many leaked passwords were successfully matched by generated passwords.

The number of matched passwords was counted (from Jaro function output) and the accuracy percentage was calculated with the matching algorithm, denoted as $A$, calculated as:

\begin{equation} \label{eq:accuracy_sym}
A = \frac{M}{N_{test}}
\end{equation}

where:
\begin{itemize}
    \item $M$ represents the total number of successful matches;
    \item $N_{test}$ represents the total number of passwords in the test set.
\end{itemize}

This measured how well the generated passwords represented real passwords.

A higher accuracy indicates:
\begin{itemize}
    \item The generated passwords were more realistic;
    \item The model better represented real user behaviour.
\end{itemize}

\subsection{Password Experiment}

For the Indian password experiment, ChatGPT generated $6,666$ Indian passwords were used. These passwords were compared with $7,675$ leaked Indian passwords. After some preliminiary experiments, Jaro similarity with a threshold of $0.5$ was found to be appropriate for usage. 



These passwords were compared with the English LinkedIn leaked password dataset using the Jaro similarity threshold of $0.5$. The number of matched passwords was recorded and converted into an accuracy percentage. This experiment allowed a comparison between training with English-only and multilingual (mixed English-Indian) passwords.


\subsection{Key Difference from Original Approach}

The original PassGAN by \cite{hitaj} and others \cite{yu2022} can be used to generate huge password datasets and analyse password frequencies. However, training such a complex neural architecture requires a high computational cost and it is shown here that it is not required as generative AI tools such as ChatGPT produces similar and realistic password list to PassGAN. Moreover, the original PassGAN model was only trained on leaked English passwords, however the approach discussed here also includes Indian passwords justifying how ChatGPT can also be used for multilingual password strength guess estimation.




\begin{figure*}[htbp]
    \centering
    \begin{subfigure}[t]{1\linewidth}
        \centering
        \includegraphics[width=\linewidth]{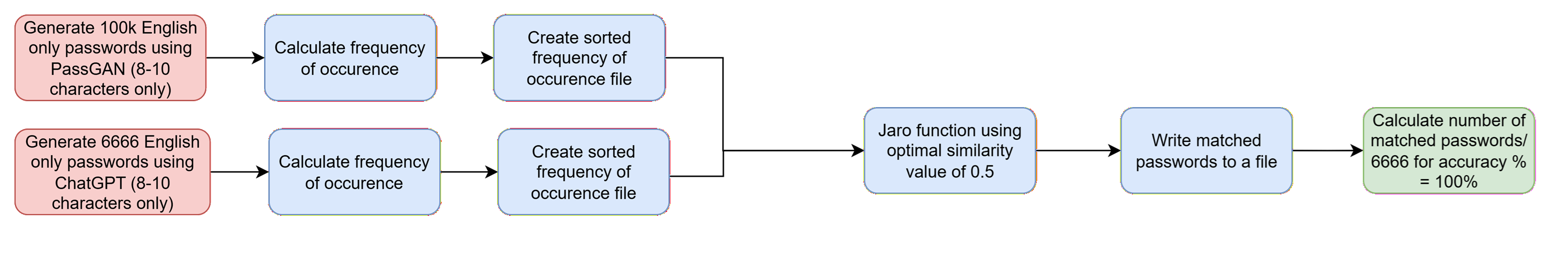}
        \caption{} 
        \label{fig:image4}
    \end{subfigure}\hfill 
    \begin{subfigure}[t]{1\linewidth}
        \centering
        \includegraphics[width=\linewidth]{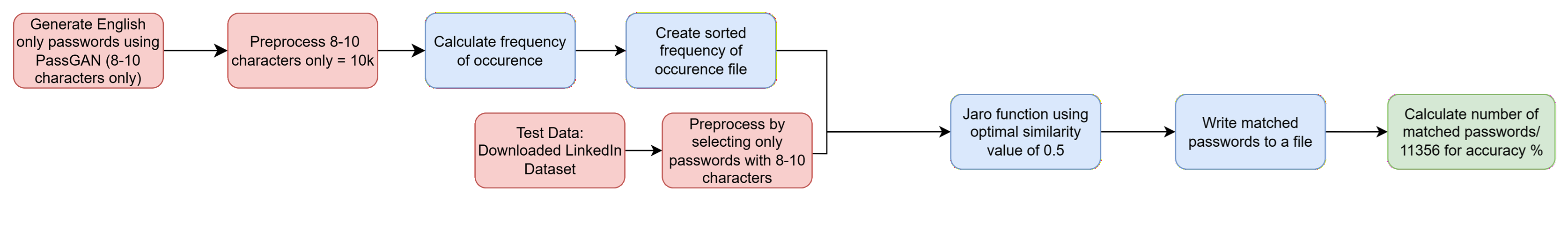}
        \caption{}
        \label{fig:image3}
    \end{subfigure}\hfill 
    \\
    \begin{subfigure}[t]{1\linewidth}
        \centering
        \includegraphics[width=\linewidth]{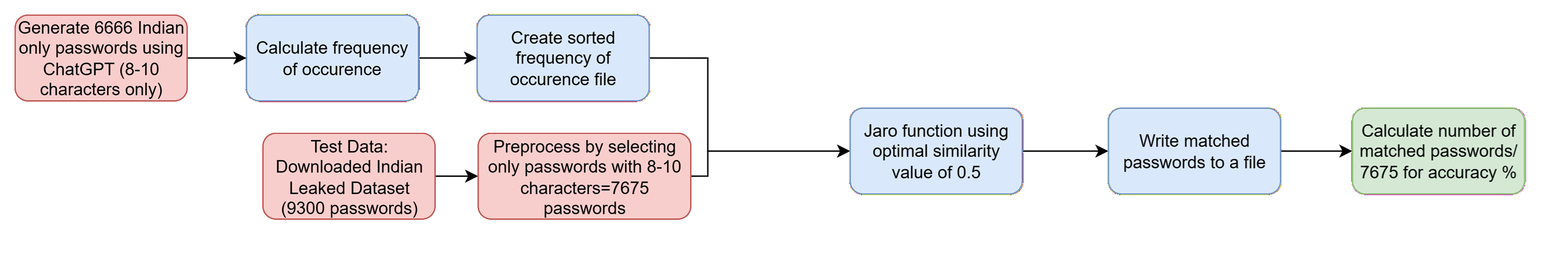}
        \caption{}
        \label{fig:image3}
    \end{subfigure}\hfill 
    \begin{subfigure}[t]{1\linewidth}
        \centering
        \includegraphics[width=\linewidth]{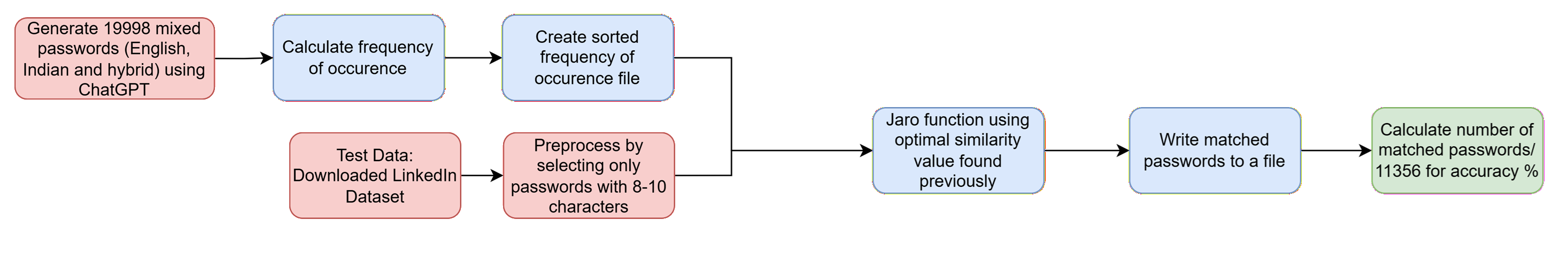}
        \caption{}
        \label{fig:image1}
    \end{subfigure}
    
    \caption{Models used in this paper: (a) English password model (PassGAN vs ChatGPT English) (b) Baseline (PassGAN) and LinkedIn test (c) Indian password model (ChatGPT vs Leaked Indian)  (d)  Mixed password model (ChatGPT vs LinkedIn)}
    \label{fig:all_images}
\end{figure*}

Figure \ref{fig:all_images} shows the flow of the processes used in the work here. A summary of the methods:
\begin{itemize}
    \item Generating password datasets using ChatGPT
    \item Using real leaked datasets for testing
    \item Matching passwords using Jaro similarity
    \item Calculating accuracy percentages for English, Indian, and mixed datasets
\end{itemize}

The approach used here allowed evaluation of whether ChatGPT could be used as an alternative to GAN-based password models and whether multilingual password generation improves performance.

\section{Results}

This section presents the results of the password matching experiments. The generated passwords were compared with real leaked password datasets using Jaro similarity.
The aim was to measure how well the generated passwords matched real-world passwords.

\subsection{PassGAN and ChatGPT Generated Matching Results}
For the English password experiment, using $6,666$ generated passwords by ChatGPT and testing of $100,000$ generated English passwords from original PassGAN model \cite{D3vil0p3r}.
Using $0.5$ Jaro similarity value, the matching was $100$\%  showing how well English-only password generation matched passwords generated from PassGAN, justifying that ChatGPT can be used instead.

\subsection{Baseline (PassGAN) and LinkedIn Test Results}
PassGAN approach from \cite{D3vil0p3r} was used to generate $10,000$ English passwords and using Jaro similiarity threshold of $0.5$ was matched with the LinkedIn test dataset, which gave 96.00\% accuracy.

\subsection{Indian-only Password Results}
In the Indian password experiment, $6,666$ generated passwords by ChatGPT were used with $7,675$ leaked Indian passwords were used as the test dataset. After applying Jaro similarity matching of $0.5$, the number of matched passwords was number of $7,673$, resulting in accuracy of $99.97$\%. Given the near perfect accuracy, other Jaro thresholds were not tested. This result shows how well the generated Indian-style passwords represented real Indian passwords.

\subsection{Mixed Language Password Results}
In the mixed language password experiment, generated passwords from ChatGPT contained English, Indian, and both English and Indian (mixed) passwords. The same leaked English LinkedIn dataset was used as the test dataset.

When looking at Table \ref{research_results}, it can be seen how well multilingual ChatGPT password generation matched real English passwords and performing better than PassGAN for a similarity threshold of $0.5$ (this was used in the Indian password set and also that gave perfect PassGAN vs ChatGPT English match). 


\begin{table}[htbp]
\centering
\caption{ChatGPT and PassGAN Results using Jaro threshold of $0.5$ against English LinkedIn Test Dataset}
\label{research_results}
\setlength{\extrarowheight}{5pt}
\begin{tabular}{|c|c|c|c|c|}
\hline
\multicolumn{4}{|c|}{\textbf{ChatGPT Generated}} & \multirow{2}{*}{\makecell{\textbf{PassGAN} \\ \textbf{(Baseline)}}} \\ \cline{1-4}
\textbf{English} & \textbf{Indian} & \textbf{Mixed} & \textbf{Total} & \\ \hline
78.08\% & 61.85\% & 30.29\% & 99.92\% & 96.00\% \\ \hline
\end{tabular}
\end{table}



\section{Evaluation}\label{AA}
This section discusses the significance of the results, summarising the key contributions and limitations of the proposed approaches. 

The results show that ChatGPT-generated passwords were able to match real leaked passwords. This is a significant result as it denotes that the generated passwords are realistic and similar to passwords created by real users and computationally complicated models like PassGAN (and leaked password databases with their ethical concerns) are no longer necessary. 

The mixed-language generated dataset achieved an accuracy of $99.92$\% with the English-only generated dataset achieved an accuracy of $78.08$\%. The PassGAN approach achieved $96.00$\%. The results show that the mixed language dataset performed better, suggesting that including additional language(s) should help model real password behaviour. This could be because of users including words from different languages when creating passwords.

The main contribution is the use of ChatGPT approach, which has several advantages:
\begin{itemize}
    \item Faster development (simple method) - the method is easy to implement and does not require complex machine learning training.
    \item Easier dataset creation (not requiring leaked database, with ethical concerns).
    \item Realistic password generation: The passwords contained real word patterns instead of random characters. This made them more realistic.
    \item Multilingual support: The system could generate passwords using multiple languages.
\end{itemize}

As such, the use of ChatGPT, although much simpler than PassGAN, especilly for multilingual passwords generation and additional training), the results show it can still generate realistic passwords. This suggests that such generative machine learning models are useful for password research.

This study is also the first to look into Indian PSM and has showed the ChatGPT based approach can be utilised for PSM for Indian passwords. Finally, the work has shown that using a mixture of languages provide a better modelling of the PSM.  


The proposed approaches do have limitations:

\textit{Limited dataset size} - the generated datasets were smaller than typical GAN utilised datasets. PassGAN studies often use millions of passwords. This study used only a small number of $6,666$ generated passwords each for English, Indian, and mixed generated datasets as there was a limit on the number of passwords ChatGPT could generate. 


\textit{Structure restrictions} - All generated passwords followed the similar structure. Real passwords can have many different structures. For this study, only passwords with lengths between $8$-$10$ characters, had at least one number, one symbol, one lowercase and one uppercase letter for both test and generated datasets (this filtering was done to simplify the experiments). 


\section{Conclusion and Future Work}
This project investigated whether ChatGPT could be used to generate realistic passwords and whether multilingual password generation improves password modelling. Three datasets were created: English, Indian, and mixed passwords. The generated passwords were compared with real leaked password datasets using similarity matching as using similarity matching allowed detection of similar passwords instead of exact matches, which better reflects real password attacks.


This research showed that ChatGPT can be used to generate realistic password datasets instead of PassGAN. The results suggest that multilingual password generation may improve password modelling (e.g., learning more cultural words) rather than a single language model. This approach provides a simple alternative to complex password generation systems. It also successfully matched a leaked Indian password database denoting its ability to be used as Indian PSM. 


%
%

Future work could look into including using larger password datasets with more languages included, especially those different semantically such as Chinese that is highly context-dependent. This would further improve accuracy and reliability. Other similarity measures such as cosine approach and vector embeddings (similar to transformers) to improve the semantic matching could also be explored such as used in \cite{tian2025}. 



\vspace{12pt}
\color{red}

\end{document}